\documentstyle{article}

\begin{document}

\renewcommand{\thefootnote}{\fnsymbol{footnote}}
\large
\begin{flushright}
BNL-49275 \\
\end{flushright}

\begin{center}
\Large
{\bf Wave Function Evolution of a Dissipative System }\\
\vspace{1cm}
\Large

Li Hua Yu\\
National Synchrotron Light Source, Brookhaven National Laboratory, N.Y.11973\\
Chang-Pu Sun\\
Institute of Theoretical Physics, State University of New York \\
Stony Brook, N.Y.11794$^{}$\footnote{\large Permanent Address:
Department of Physics, Notheast Normal University, Changchun 130024, P.R.China}
\vspace{0.4cm}

Abstract\\
\end{center}

For a dissipative system with Ohmic friction,
we obtain a simple and exact solution for the
wave function of the system plus the bath. It is
described by the direct product in two independent Hilbert space.
One of them is described by
an effective Hamiltonian, the other represents
the effect of the bath, i.e., the Brownian motion,
thus clarifying the structure of the wave function of the system
whose energy is dissipated by its interaction with the bath.
No path integral technology is needed in this treatment.
The derivation of the Weisskopf-Wigner line width theory follows easily.

\newpage
The simplest example of a dissipative system, an harmonic oscillator
coupled to the environment, which is a bath of harmonic oscillators,
has been the subject of extensive studies$^{1-15}$.
We shall show in the present paper that in a special case,
the Ohmic case (to be defined later), the dissipative system can
be exactly treated both classically and quantum mechanically,
thereby clarifying the sense in which the wave function
is describable by an effective Hamiltonian. In this treatment
path integral technology is not needed, and our presentation
is self-contained.

We consider the problem discussed by Caldeira and Leggett (CL)$^1$,
an harmonic oscillator
system (the dissipative system) with coordinate q, mass M,
and frequency
$  ( \omega_0^2 + \Delta \omega^2 )^{1/2} $, interacting
with a bath of N harmonic oscillators of coordinates $x_j$,
mass $m_j$, and frequency $\omega_j$.,
where $\Delta \omega^2$ is a shift induced by the coupling already
discussed by CL.
The Hamiltonian of the system and the bath is:
$$H=\frac{p^2}{2M} + \frac 1 2 M
( \omega_0^2 + \Delta \omega^2 ) q^2
+ q \sum_j c_j x_j
+ \sum_j \left(  \frac {p_j^2}{2 m_j}
+ \frac {1} {2} m_j \omega_j^2 x_j^2 \right)  \; \; \; .\eqno{(1)}$$

The dynamic equation for operators in the Heisenberg representation leads
to the following set of equations of motions:
$$
M \ddot{q}= - M \omega_0^2 q + M \Delta \omega^2 q
- \sum_j c_j x_j  \; ,
\eqno{(2)}$$
$$
m_j \ddot{x} = -m_j \omega_j^2 x_j  - c_j q  \; \; \; \;  (j=1,2,....N).
\eqno{(3)}$$
Now, applying the Laplace transform$^2$
(the bars are our notations for the Laplace transform, s is the
Laplace transform of time t),
equations (2), (3) can be used to eliminate
bath variables
$\bar{x}_j$ to obtain the equation for $\bar{q}$
$$
M( s^2 \bar{q} -s q_0 - \dot{q}_0 )
= -M \omega_0^2 \bar{q} - M \Delta \omega^2  \bar{q}
- \sum_j  c_j \frac{ s x_{j0}+ \dot{x}_{j0}  }{s^2 + \omega_j^2}
- \sum_j  \,  \,  \frac{c_j^2}
{m_j (s^2 + \omega_j^2 )} \bar{q}  \, ,
\eqno{(4)}$$
where $q_0 , \, \dot{q}_0 , \, x_{j0} , \,  \dot{x}_{j0} $
are the initial values
of the respective operators in the Heisenberg representation.
Assuming the number of bath oscillators is large enough so that
we can replace the sum over j by an integration over $\omega_j$,
the coefficient of the last term can then be separated into two terms:
$$
\int_0^{\infty}  \,  \,
\frac{\rho ( \omega_j ) c_j ^2 d \omega_j }
{m_j   \,  \omega_j^2 }
- s^2   \,  \,  \int_0^{\infty}  \,
\frac{ \rho ( \omega_j  )  \,  c_j^2  d \omega_j }
{ m_j  \omega_j^2 ( s^2 + \omega_j^2 )}
 \, ,
\eqno{(5)}$$
where $\rho ( \omega_j )$ is the bath oscillator density.
Following an argument similar to the one
pointed out by CL$^1$, the requirement that the system becomes
a damped oscillator with frequency $\omega_0$ and damping rate $\eta$
in the classical limit, known as the ``Ohmic friction'' condition,
leads to the following constraint:
$$
\rho ( \omega_j ) = \frac {2 \eta M } \pi \frac{ m_j
\omega_j^2 }{ c_j^2  } .
\eqno{(6)}$$
By observing eq.(4) and eq.(5), it can be shown that the second term of
eq.(5) leads to damping with the damping constant $\eta$,
while the first term of eq.(5)
represents a frequency shift. If the frequency renormalization
constant $\Delta \omega^2$ is chosen to satisfy:
$$
M \Delta \omega^2  =
\int_0^{\infty}  \,  \,
\frac{\rho ( \omega_j ) c_j^2 d \omega_j }
{m_j  \omega_j^2 }  \, ,
\eqno{(7)}$$
the frequency is shifted to
$\omega_0$.
Then equation (4) is simplified, and
its inverse Laplace transform gives the quantum Langevin equation,
valid at time t $ >=  \,   0  +$:
$$
\ddot{q} + \eta \dot{q} + \omega_0^2 q = f(t)   \,  \,  \, ,
\eqno{(8)}$$
with the Brownian motion driving force:
$$
f(t)=- \sum_j   c_j (x_{j0} \, cos \omega_j
+ \dot{x}_{j0} \frac{sin \omega_j t}{\omega_j} )  \,  \,  \, .
\eqno{(9)}$$
During the derivation, in order to carry out the integral in eq.(5),
we used the requirement of the inverse Laplace
transform that s must pass all the singular points from right of the
complex plane, and hence real(s) $>$0.

Equation (8) and (9) are the equations of a
\it
driven damped harmonic oscillator,
\rm
the solution of which is well known as a linear combination of
the initial values at $q_0 $, $ \dot{q}_0 $,
$x_{j0} $, and $ \dot{x}_{j0} $ :
$$
q(t) = a_1 (t) q_0 + a_2 (t) \dot{q}_0
+ \sum_j  \, ( b_{j1} (t)  x_{j0}  + b_{j2}  (t) \dot{x}_{j0} )  \,  \,  \, ,
\eqno{(10)}$$
$$
x_i (t) = \alpha_{i1} (t) q_0 + \alpha_{i2} (t) \dot{q}_0
+ \sum_i  ( \beta_{ij1} (t)  x_{j0}  + \beta_{ij2}  (t) \dot{x}_{j0} )  \, , \,
 with
\eqno{(11)}$$
$$
a_1 (t) = \frac{\nu e^{- \mu t} - \mu e^{-\nu t} }
{\nu  \,  -  \,  \mu }  \, , \,
a_2 (t) = \frac { e^{- \mu t} -  e^{-\nu t} }
{\nu  \,  -  \,  \mu }  \, ,
\eqno{(12)}$$
$$
\mu=\frac{\eta} 2 + \omega  \,  i  \, ,
 \,  and  \,  \,  \nu= \frac{\eta} 2 - \omega  \,  i  \,  ,
\eqno{(13)}$$
and here $\omega = ( \omega_0^2 - \eta^2 /4 )^{1/2} $
is the frequency
shifted by  damping.
(All formulae are correct whether $\omega$ is real or imaginary.
To avoid a minor detail of the initial value problem, we
have redefined the initial time as t=$0  +$.)
The explicit expression for
$b_{j1} $, $ b_{j2} $,
$\alpha_{i1} $, $ \alpha_{i2} $, $ \beta_{ij1} $, and $ \beta_{ij2} $
are well known in freshmen physics.

We emphasize
that the use of the Laplace transform instead of the Fourier transform
allows us to express q(t) and $x_j (t)$
explicitly in terms of the initial values, as in eq.(10) and eq.(11).

The equations (10) and (11)
serve as the starting point of subsequent discussions. We will
proceed to find the Green's function of the full system, and hence
the solution of the wave function
in Schoedinger representation. The result tells us
\it
in what sense the damped oscillator is described by
an effective Hamiltonian without the bath variables
\rm
and gives its specific form, it also shows that
under this condition, the wave function can be factorized, and the main
factor relevant to the damped oscillator is a solution of the
Schoedinger equation with an effective Hamiltonian.

Equations (10) and (11) are correct both in classical mechanics and in
quantum mechanics in the Heisenberg representation. We notice that
q(t) and $x_j$(t) are both linear superpositions of
$ q_0 , \,  \dot{q}_0 , \,  x_{j0} ,  \,  \dot{x}_{j0} $ with c-number
coefficients. The commutation rules between
$ q (t) , \,  \dot{q} (t) , \,  x_j (t) ,  \,  \dot{x}_j (t) $ are
$[q(t), \dot{q} (t)] \, =\frac{  i \hbar  } M  \, ,
 \,  [ x_j (t) ,  \,  \dot{x}_j (t)] \, =\frac{ i \hbar }{m_j} $,
and operators q(t) and $\dot{q} (t)$ commute with
$ \,  x_j (t) $ and $ \dot{x}_j (t) $.
One can prove these commutation rules
by two ways: (a). By direct computation,
using the fact that at t=0, they are correct.
(b). By the general principle that
$ q (t) , \,  \dot{q} (t) , \,  x_j (t) ,  \,  \dot{x}_j (t) $
are related by a unitary transformation to
$ q_0 , \,  \dot{q}_0 , \,  x_{j0} ,  \, and  \,  \dot{x}_{j0} $.

Equation (10) and (11) show that the operators q(t) and $x_j (t)$
can each be written as a sum of two terms:
$$
q(t)= Q(t) + \sum_j  \,   \xi_j (t)  \,  \, ,
\eqno{(14)}$$
$$
x_i (t) = \zeta_i (t) + \sum_j  \,  X_{ij} (t)  \,  \,  \, ,
\eqno{(15)}$$
where $ Q(t)$ and $\zeta_i (t)$ are linear in $q_0 $ and $\dot{q}_0$
and independent of $ x_{j0}  $ and $  \dot{x}_{j0}$, and $\xi_j$ and
$X_{ij} (t)$ are linear in $x_{j0}  $ and $  \dot{x}_{j0} $, and
independent of $q_0  $ and $  \dot{q}_0 $. Thus
Q(t) and $\zeta_i (t)$ are operators in
one Hilbert space $ S_Q$, while $ \xi_j (t)$ and
$X_{ij} (t)$ are in an independent Hilbert space $S_X$,
and the full Hilbert space is a direct product $S_Q
\bigotimes
S_X$.

We shall first analyze the structure of $S_Q$. We write that:
$$
Q(t) = a_1 Q_0 + a_2 \dot{Q}_0 =
 a_1 Q_0 - a_2  \, \frac{i \hbar}M   \frac{\partial}
{ \partial  Q_0 }  \,  \,  \, .
\eqno{(16)}$$
To explicitly show that we are discussing the $S_Q$
space, we define
$Q_0 \equiv q_0 $, $ \dot{Q}_0 \equiv \dot{q}_0$.
The eigenfunction of Q(t) with an eigenvalue denoted by $Q_1$,
in the $Q_0$ representation, is easily calculated to be:
$$
u_{Q_1} ( Q_0  , t)=
\left( \frac{M \omega e^{ \frac 1 2 \eta t } }{2 \pi \hbar sin \omega t }
\right)^{ \frac1 2}
exp \left[ - \frac{iM}{ 2 \hbar a_2}
\left( a_1 Q_0^2 - 2 Q_0 Q_1 +
\phi  ( Q_1 , t) \right) \right]  \; \; ,
\eqno{(17)}$$
with $\phi $ as an arbitrary phase, i.e., a real number.
This eigenfunction
is related to the Green's function $G(Q_1 ,  Q_0  ;  t ,  0 )
 \equiv   < Q_1 | U_Q  (t) | Q_0 > $, by
$u_{Q_1} ( Q_0  , t)= < Q_0 | U_Q^{-1}
(t) | Q_1 >$ $ = < Q_0 | U_Q ^{\dagger}
(t) | Q_1 > =G^* (Q_1 ,  Q_0 ;  t ,   0 )$,
where we denote the evolution operator by $ U (t)$, which is unitary
when we choose the eigenvectors of Q(t) to be orthonormal.
Thus we have:
$$
G ( Q_1 , Q_0 ;t,0)=
\left( \frac{M \omega e^{\frac1 2 \eta t } }{2 \pi  \hbar  sin \omega t }
\right)^{\frac1 2}
exp \left[ \frac {iM}{ 2 \hbar a_2}
\left( a_1 Q_0^2 - 2 Q_0 Q_1 +
\phi  ( Q_1    ,   t) \right) \right]  \; \; .
\eqno{(18)}$$

Next, we shall determine
the arbitrary phase $\phi (Q_1 , t )$, which is the phase
of the eigenvectors of Q(t).
Using eq.(16), we find the commutation rule
for $Q$ and $\dot{Q}$:
$$
[ Q,     \dot{Q} ] = ( a_1 \dot{a}_2 - \dot{a}_1 a_2 )
[Q_0 ,     \dot{Q}_0 ] =
e^{- \eta t} [ Q_0,    \dot{Q}_0 ] =\frac { e^{ - \eta t } }
M i \hbar  \,   \, .
\eqno{(19)}$$
Thus we define the canonical momentum P(t) as:
$$
P(t)=M   e^{  \eta t } \dot{Q} (t)
=M~e^{  \eta t }
( \dot{a}_1 Q_0 + \dot{a}_2 \dot{Q}_0 )  =
 \dot{a}_1 Q_0 - \dot{a}_2 \frac{i \hbar} M  \frac{\partial  }
{ \partial  Q_0 }     \,  \, ,
\eqno{(20)}$$
and get the commutation rule: $ [Q(t),P(t)]=i \hbar     $.
The eigenfunction of P(t) can be calculated in two ways:  (a).
We can calculate the eigenvector of P(t) in the $Q_0$ representation
using eq. (20) and then use the Green's function eq.(18) to transform it
into the Q(t) representation;  (b). The commutation rule
$ [Q(t),P(t)]=i \hbar    $ requires
that P(t)=$ -i \hbar  \frac{ \partial  \,  \,  \, }{\partial Q} $,
so the eigenfunction of P(t) with
eigenvalue $P_1$ is $exp [ i \frac{P_1}{\hbar } Q  ]$.
By comparing these two solutions, the arbitrary phase $\phi
(Q_1 ,   t)$ in the Green's function
is determined to within a phase $\phi (t)$, which is
independent of $Q_1$.
$\phi (t)$ is an arbitrary real function of time,
except that $\phi (0)=0$ so that it satisfies the condition that at t=0,
the Green's function becomes $\delta ( Q_1 - Q_0 )$.
Thus we obtain the Green's function in the $S_Q$ space:
$$
G(Q_1 , Q_0 ;t,0)=
\left( \frac{M \omega e^{\frac1 2 \eta t }}{2 \pi i \hbar sin \omega t }
\right)^{\frac1 2}
exp \left[ \frac{iM}{ 2 \hbar      a_2}
\left( a_1  Q_0^2 + \dot{a}_2 e^{ \eta t } Q_1^2
- 2 Q_0 Q_1  \right) - \frac{i \phi (t)} { \hbar }    \right] .
\eqno{(21)}$$

It is then straight forward to derive the Hamiltonian $H_Q$
using the following relation:
$$
H_Q = i \hbar \frac{ \partial U_Q }
{ \partial t} U_Q^{-1}  \, ,
\eqno{(22)}$$
and remembering that the matrix elements of $U_Q$
and $U_Q^{-1}$ are the Green's function and its conjugate.
The result is:
$$
H_Q =  e^{ - \eta t}  \frac{P^2}{2 M}
+\frac1 2 M \omega_0^2     e^{\eta t}      Q^2
+  \dot{\phi} (t)  \,  \,  \,  \, .
\eqno{(23)}$$
Since $\phi$ is arbitrary except that  $\phi  (0) =0$, we can take
$\phi (t)=0$. Therefore we have derived the well known
effective Hamiltonian for the dissipative system.
\it
We emphasize that the expression for $ H_{Q}$
is here derived, while in
usual literature it is introduced by more or less heuristic arguments.
\rm

Next, we shall analyze the effect of the bath. Similar to
eq. (17) we obtain the eigenfunctions
$\theta_{{\xi}_{j1}} (x_{j0}    ,   t) $ for $\xi_j$.
Using Dirac's notation we have:
$ Q |u_{Q_1} > = Q_1 |u_{Q_1} > $,
$\xi_j | \theta_{{\xi}_{j1}} > = \xi_{j1}    | \theta_{{\xi}_{j1}} >$.
Thus $\displaystyle | u_Q >     \prod_j
\bigotimes |  \theta_{\xi_j} >$ is
an eigenvector of q(t), with eigenvalue of $\displaystyle Q + \sum_j
\xi_j$. In other words, the eigenvector of q(t) with eigenvalue q
is
$\displaystyle |q, \{  \xi_j   \} > = |u_{\displaystyle q- \sum_j \xi_j}  >
\prod_j \bigotimes | \theta_{\xi_j}  > $.

If initially the wave function is
$\displaystyle | \Psi_0 > = |  \psi_0 > \prod_j \bigotimes
 | \chi_{j0}  >$,
to calculate the wave function at time t, we should expand $\Psi_0$ in terms of
the eigenvectors $|q, \{  \xi_j   \} >$, i.e., we should calculate
$$
\psi ( Q ,   t)    \equiv    < u_ Q | \psi_0 >
= \int u_Q^* ( q_0 )  \psi_0 ( q_0 )    d q_0     \, ,
\eqno{(24)}$$
$$
\chi_j ( \xi_j ,     t)    \equiv   < \theta_{\xi_j} | \chi_{j0} > =
\int \theta_{\xi_j}^*
( x_{j0} )     \chi_{j0} ( x_{j0} )    d x_{j0}  \,  \, .
\eqno{(25)}$$
Then the wave function in the Schoedinger representation at time t is
$$
\Psi ( q, \{ \xi_j \}    , t) = < q, \{ \xi_j \}| \Psi_0 > =
\psi ( q     -    \sum_j \xi_j    ,   t)  \,
\prod_j     \chi_j ( \xi_j ,     t)  \,  \, .
\eqno{(26)}$$

Notice that $\psi (Q,t)$ of eq.(24)
is the wave function in the Schoedinger representation
with the effective Hamiltonian eq.(23). Hence we have connected the
effective Hamiltonian approach to the dissipative system problem
with the other approaches that take both the system and the bath into
account. We also notice that even though our
$\Psi ( q, \{ \xi_j \} , t )$ is
in a different representation from that of $\Psi ( q, \{ x_j \} , t )$,
the usual probility interpretation is still valid:
$\displaystyle \int \int ... \int
| \Psi ( q, \{ \xi_j \} , t ) |^2 \prod_j    d    \xi_j$ is
the probability density of finding the particle at q.
Since
this solution is very simple, it provides a simple way to analyze other
complicated problems, e.g.,
study the influence of Brownian motion on interference,
which we shall not elaborate because of space.

Under certain conditions, eg., at low temperature and when the system q
is in highly excited states so the range of q is large enough
that for all the values of $\xi_j$ which do not have vanishingly small
probability,
$q \gg \displaystyle | \sum_j     \xi_j |$, we can approximately write:
$\displaystyle  \Psi ( q,     \{ \xi_j \}    , t) =
\psi ( q    ,   t)     \prod_j     \chi_j ( \xi_j ,     t) $.
That is, the wave function is factorized, the dissipative system q
can be described by the wave function $\psi (q,    t)$ only, and
the Brownian motion can be ignored.
Therefore it is interesting to examine the width of the
argument of the wave function $\psi$, due to the Brownian motion, i.e.,
the mean value of $\displaystyle ( \sum_j \xi_j )^2 $
at time t. It can be calculated using its expression in the Heisenberg
representation
as introduced by (10) and (18). At temperature T, this width is :
$$
< ( \sum_j \xi_j  (t) )^2 > =
\sum_j  \frac{ \hbar } {2 m_j \omega_j }
( | b_{j1} (t) |^2 + \omega_j^2 | b_{j2} (t)  |^2 )
\, cot \left( \frac{ { \hbar } \omega_j }{ 2kT} \right)    \,  \,  \, .
\eqno{(27)}$$
This width is zero initially, then approaches its final equilibrium value
in a time interval of the order of 1/$\eta$.
At low temperature limit, the equilibrium width is simplified to:
$$
\sigma^2 ( t = \infty ) =\frac{\hbar}{2 \pi m \omega } \left[\frac\pi 2 +
arctan \left( \frac{\omega_0^2  }{ \eta \omega } \right) \right ]  \,  \, .
\eqno{(28)}$$
If the damping rate $\eta$ is much smaller than the frequency of the
oscillator $\omega_0$, this width happens to become the same as
the width of the ground state of the system $ {\hbar }/(2m \omega )$.

Finally, it is interesting to see the distribution of the dissipated
energy of the harmonic oscillator in the bath, and check if it agrees
with the Weisskopf-Wigner line width theory$^{16}$.
For simplicity, we assume zero temperature.
To calculate the energy dissipated by the system into the j'th
bath oscillator, we use eq.(11) and its derivative to obtain the
expression of $x_j (t)$ and $    p_j = m_j \dot{x}_j (t) $,
which are
then substituted into the expression of the energy of the j'th oscillator:
$ h_j = \frac{p_j^2}{2 m_j}
+ \frac1 2 m_j \omega_j^2 x_j^2   $.
We then calculate the expectation value of $h_j$,
assuming initially the system is in the n'th excited state,
and the bath oscillators in the ground state.
We calculate the contribution to the expectation value
of $h_j$ from the system, by keeping only terms which depend
on $q_0$, and $ \dot{q}_0$. The result is then the
energy disspation by the system into the j'th oscillator.
When multiplied by the density of states $\rho ( \omega_j ) $ eq.(6),
it gives the dissipated energy spectrum.
It is a function of $\omega_j$, with a narrow peak near
the resonance $\omega_j = \omega $, if the damping is small,
i.e., if $\eta \ll \omega$. It is oscillatory with a frequency
of $2 \omega_j$. Its time average over a period is :
$$
E_j = (n     + \frac1 2 ) \hbar \omega \,    \frac{\eta}\pi  \frac1
{ \frac{ \eta^2 } 4 + ( \omega - \omega_j )^2 }     A ( \omega_j )   ,
\eqno{(29)}$$
where $A ( \omega_j )$ varies slowly near the resonance, i.e.,
over the width $ \eta / 2$ of the peak it changes very little.
Therefore, we can replace $\omega_j$ by $\omega$ near the peak,
and the result is simplified to:
$ A ( \omega )     \approx    1$.
Thus eq.(29) shows that the dissipated energy has a Lorenzian
distribution near the resonance, in agreement with the
Weisskopf-Wigner line width theory.

\begin{center}
{\bf Acknowledgements}
\end{center}

We thank Prof.C.N. Yang for drawing our attention to the
problem of dissipative systems, for spending his
valuable time in many sessions of
stimulating discussions on this subject, and for many suggestions
which are critically important for the ideas of this paper.
The part of work by Li Hua Yu is performed under the auspices of the U.S.
Department of Energy under Contract No. DE-AC02-76CH00016.
The part of work by Chang-Pu Sun is supported in part by a
Cha Chi Ming fellowship
through the CEEC program at the State University of New York at Stony
Brook, and in part by the NSF of China through the Northeast Normal University.

\newpage
\begin{center}
{\bf References}
\end{center}
\vspace{0.5cm}
\begin{enumerate}

\item
A.O. Caldeira, A.J. Leggett, Ann. Phys. 149, 374 (1983), and
Physica 121A, 587 (1983)
\item
A.O. Caldeira, Helvetica Phisica Acta, 61, 611 (1988)
\item
E. Kanai, Prog. Theor. Phys. 3, 440 (1948)
\item
S. Nakajima, Prog. Theor. Phys. 20, 948 (1958)
\item
R.Zwanzig, J. Chem. Phys. 33, 1338 (1960)
\item
I.R. Senitzky, Phys. Rev. 119, 670 (1960)
\item
R.P. Feynman, F.L. Vernon, Ann. Phys. 24, 118 (1963)
\item
G.W. Ford, M. Kac, P. Mazur, Jour. Math. Phys., 6, 504 (1965)
\item
M. D. Kostin, J. Chem. Phys. 57, 3589 (1972)
\item
W.H. Louisell, "Quantum Statistical Properties of Radiation",
John Wiley \& Sons (1973)
\item
M. Sargent III, M.O. Scully, W.E. Lamb, "Laser Physics", Addison-Wesley (1974)
\item
K. Yasue, Ann. Phys. 114, 479 (1978)
\item
R.H. Koch, D.J. Van Harlingen, J. Clarke, Phys. Rev. Lett, 45, 2132 (1980)
\item
H. Dekker, Phys. Report, 80, 1 (1981)
\item
R. Benguria, M. Kac, Phys. Rev. Lett, 46, 1 (1981)
\item
V.F. Weisskopf,  F.P. Wigner, Z. Physik 63, 54 (1930); 65, 18 (1930)

\end{enumerate}
\end{document}